# Comments on 'Revisiting building block ordering of long-period stacking ordered structures in Mg-Y-Al alloys'

Xin-Fu Gu[a, b*], Tadashi Furuhara[b], Leng Chen[a], Ping Yang[a]

[a] School of Materials Science and Engineering, University of Science and Technology Beijing, Beijing, 100083, China

[b] Institute for Materials Research, Tohoku University, Sendai, 980-8577, Japan

*Corresponding author: Xin-Fu Gu, xinfugu@ustb.edu.cn

**Abstract**

In a recent paper, Zhang *et al.* [Acta Materialia 152 (2018) 96] studied the in-plane ordering of the long-period stacking ordered (LPSO) structure in Mg-Al-Y alloys. In addition to the well-known $L1_2$ type building cluster, they proposed three new types of metastable building clusters. However, we will show that these new types of building clusters are caused by the superimposition of $L1_2$ type clusters located in different domains. In addition, the experimental evidence for domain structures in a similar alloy system Mg-Al-Gd is provided.

**Keywords:** Magnesium alloy; LPSO; Domain; HAADF-STEM

In a recent work, Zhang *et al.* studied the building block ordering of the long-period stacking ordered (LPSO) structure in Mg-Al-Y alloy [1]. As a main conclusion, they claimed that at least three new types of metastable building clusters enriched of Al and Y atoms exist in the as-casted alloy in addition to a well-known L1$_2$ (Al$_6$Y$_8$) cluster based on their observations by high angle annular dark field (HAADF)-scanning transmission electron microscopy (STEM). These three metastable building clusters will be transformed to the stable L1$_2$ cluster after further annealing at 530-550°C, and this evolution was validated by first-principles calculation. However, it will be shown here that these new clusters in their work could be rationalized by overlapped L1$_2$ clusters.

The four-layers height fcc structural unit (SU for short hereafter) is a building block for the LPSO structures [2]. The fcc SUs in a LPSO structure are separated by several (0001)$_\alpha$ Mg (hcp) layers, and the separation varies with different types of LPSO structures. Figure 1(a) shows the projection of ideal SU along [0001]$_\alpha$ in Mg matrix, where the SU enriches with rare earth (RE) and metallic elements (M) [3, 4] and contains ordered distributed M$_6$RE$_8$ clusters with L1$_2$ type structure [4, 5]. The L1$_2$ cluster in Figure 1(a) is shown in Figure 1(b). The atoms in "ABCA" stacking layers are denoted by circles, downward-pointing triangle, upward-pointing triangle, and circles, respectively. The <111> view of the L1$_2$ structure forms a cluster in Figure 1(a), the neighboring clusters in a rhombic lattice are separated by $2\sqrt{3}a$ along <10-10>$_\alpha$, where *a* is lattice parameter of Mg matrix. Figure 2(c) and (d) show the <11-20>$_\alpha$ and <10-10>$_\alpha$ view of the SU, respectively. The corresponding HAADF-STEM image taken from our recent studied Mg-3Al-5Gd alloy [6] is shown at the right side of each atomic model for reference (same for the rest). The bright dots in the images indicates the locations of Gd enriched columns according to Z-contrast principle [7]. The ordered pattern in Figure 2(d) is denoted as "R1" type cluster according to Ref. [1].

For the L1$_2$ clusters in Figure 1(a), if we move the clusters within the dashed hexagon relatively to their original positions, a domain structure will be generated.

The possible translation positions could be classified into four types as indicated by different symbols in the rhombic unit cell in Figure 2(a). The nodes of the grid in the figure are the positions of Mg atoms in stacking layer A. The original position for Gd atoms in the clusters in layer A is denoted as $t_1$ position, and translation to such positions will not generate domain boundaries. Therefore, there are three crystallographic inequivalent positions to generate a domain structure, i.e. $t_2$, $t_3$ and $t_4$. Due to the overlap of the domain structures, the $<11\text{-}20>_\alpha$ and $<10\text{-}10>_\alpha$ view of the SU will be different from Figure 1(b-c). Figure 2(b) shows an example of domain structure with a translation of $t = t_4 = a$. This kind of translation will result in two types of $<11\text{-}20>_\alpha$ view of Figure 2(b) as shown in Figure 2(c), and two types of $<10\text{-}10>_\alpha$ view as shown in Figure 2(d). The ordered patterns in Figure 2(d) are denoted as "R4" and "R2" in Ref. [1], respectively. The formation of "R4" and "R2" is due to the overlap of the $L1_2$ clusters in two domains. Moreover, the cluster "R3" in Ref. [1] can be obtained from a $<10\text{-}10>_\alpha$ view with translation to $t_3$ positions. Table 1 summarizes the possible $<10\text{-}10>_\alpha$ view of a SU contain a domain structure with the inequivalent translation positions specified in Figure 2(a), and the observed cluster R1~R4 is marked with "√" in the table. It appears possible to explain all of the four clusters R1~R4 in Ref. [1] based on the overlap of $L1_2$ clusters due to the domain structures. Indeed, the domain structures in LPSO structure have been directly observed by Scanning Tunneling Microscopy [8, 9]. In considering of the focal depth during HAADF-STEM imaging, the overlap of domain structures is possible. If several domain structures are overlapped, more complex ordered patterns could be observed. It should be noted that the present analysis is pure geometry, and the relaxed structures may vary a little from the geometrical models.

The tilted views of the SU containing domains in a similar alloy system Mg-3Al-5Gd are also consistent with above analysis. The LPSO or SU usually has a large aspect ratio, and it is hard to find the same position during tilting of zone axis. However, there are steps during growth of LPSO structure [10], and these steps can be used as a mark for finding the same position during the tilted trails. Figure 3 shows such an example. Figure 3(a) and (b) show the LPSO structure at $[11\text{-}20]_\alpha$ and $[01\text{-}10]_\alpha$ zone axis, respectively. With the aid of the steps as a marker, the nearly the same positions could be located in Figure 3. In the area marked with dashed box, the ordered pattern in Figure 3(b) is "R2" type, and the possible translation position for the domain structure is $t_2$ or $t_4$ according to Table 1. Based on Figure 3(a), the translation position could be further narrow to be $t_4$, since the $<11\text{-}20>_\alpha$ view of $t_2$ domain is the same as Figure 1(c) and different from Figure 3(a). Besides, Figure 3(a) is similar to the second atomic model in Figure 2(c) with $t_4$ domain. It shows that our analysis is self-consistent, and further experimental results are also consistent using similar analysis.

Figure 4 shows the domain structure in SU viewed from $[0001]_\alpha$ zone axis. The distributions of the clusters in three areas are indicated by grid lattices. As is noted, these three lattices deviate from each other and form domains of clusters in local areas. These domain structures may happen during ordering of the planar segregation [6, 11-13]. Nevertheless, the direct observation of domain structures further supported our argument that the new clusters in Ref. [1] are possibly caused by overlapped $L1_2$ structures.

Although Zhang *et al.* considered the overlap of $L1_2$ structures, they rule out this possibility. The deficiencies in their analysis are as follows. Firstly, three possible shift vectors $r_1 \sim r_3$ are considered in their study, but the translational vectors $r_1 \sim r_3$ are crystallographically equivalent, and equal to $t_4$ in our study. However, the projections in Figures 8(f-h) in Ref. [1] are different, and only Figure 8(f) is consistent with our study. Secondly, the vectors $r_1 \sim r_3$ are the translational vectors in basal plane, and it is unreasonable that the projections of atoms in $<11\text{-}20>_\alpha$ directions will create extra positions as in subplots ③ and ④ in Figure 8(g) and 8(h) in Ref. [1]. Finally, as shown in Figure 1 and 2, our clearer HAADF-STEM images are in reasonable good agreement with the atomic models despite of the small deviation possibly due to the atomic relaxation. Moreover, the evolution of R2~R4 clusters to R1 may be due to the growth of the domain structures and the elimination of the domain boundaries.

In summary, our analysis shows that three new clusters observed in Ref. [1] may be possibly due to the overlap of the well-known $L1_2$ clusters in different domains. There are four crystallographically inequivalent translational positions to generate the domain structure. The analysis of overlapped $L1_2$ clusters in Ref. [1] is incomplete, and only one type of translation position is considered. Furthermore, our analysis is supported by the HAADF-STEM images at different zone axis in a similar alloy system Mg-Al-Gd, and the domain structure in this alloy is directly observed.

**Acknowledgement**

This work was supported by a Grant-in-Aid for Scientific Research on Innovative Areas, "Synchronized Long-Period Stacking Ordered Structure", from the Ministry of Education, Culture, Sports, Science and Technology, Japan (No.23109006) and the Fundamental Research Funds for the Central Universities (No. FRF-TP-17-003A1). The Mg alloy used in this study was provided by Kumamoto University. Special thanks to Mr. Y. Hayasaka at Electron Microscopy Center (Tohoku University) for technical supports.

Tables

Table 1. The possible cluster structures ($R_1$~$R_4$) observed along $<10\text{-}10>_\alpha$ zone axis for different domain structure with its translation vector $t$ defined in Figure 2a. The symbol "√" indicates corresponding structure would be observed, while "×" indicates the structure will not appear.

| $t$ | $R_1$ | $R_2$ | $R_3$ | $R_4$ |
|---|---|---|---|---|
| $t_1$ | √ | × | × | × |
| $t_2$ | × | √ | × | × |
| $t_3$ | √ | × | √ | × |
| $t_4$ | × | √ | × | √ |

Figures

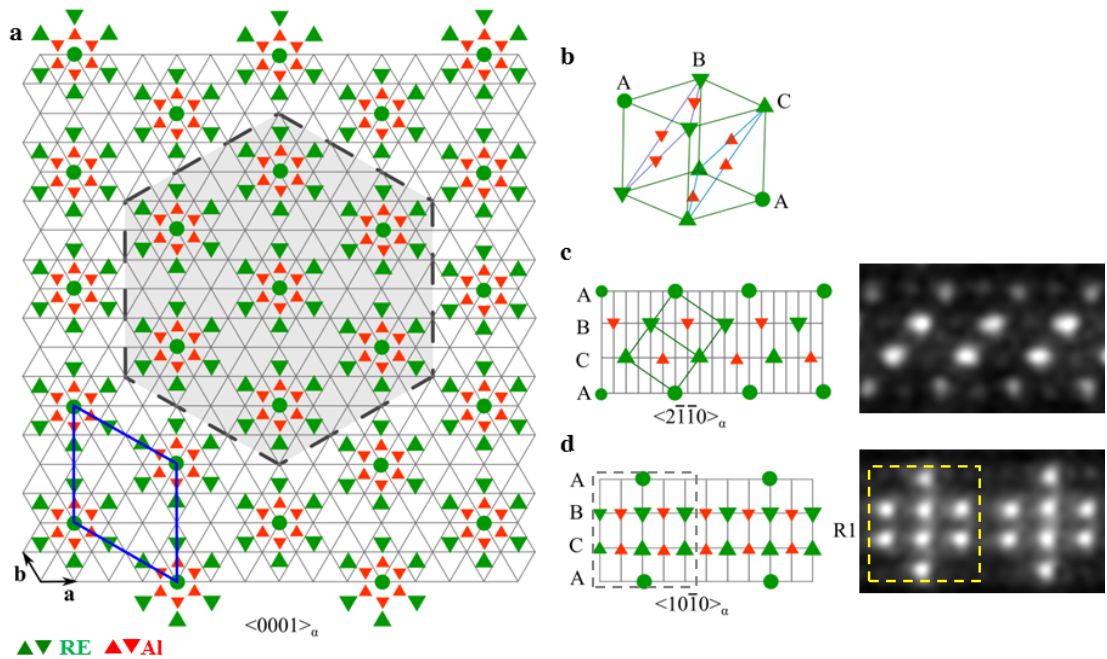

Figure 1. Schematic diagram of in-plan ordered clusters distributed in fcc structure unit of LPSO structure. a) The four-layer fcc structure unit with stacking sequence of "ABCA" viewed along [0001]$_\alpha$ in Mg ($\alpha$) matrix. Green color represents rare earth (RE) atoms, while red color is for Al atoms. The filled circles indicate the atoms in stacking layer A, upward-pointing triangle is for atoms in layer B, and downward-pointing triangle is for atoms in layer C. The nodes of grey grid show the Mg position in basal plane (A layer), and Mg atoms are omitted for simplicity. b) The L1$_2$ structure of the cluster in a). c) The <11-20>$_\alpha$ view of the fcc structure unit. A corresponding HAADF-STEM image is attached at right side for reference, where the brightest positions are location of RE atoms due to Z contrast principle. d) The <10-10>$_\alpha$ view of the fcc structure unit. The L1$_2$ cluster in d) is marked by a dashed box. This ordered pattern is named as "R1" type structure in Ref. [1]. (For interpretation of the references to color in this figure legend, the reader is referred to the web version of this article.)

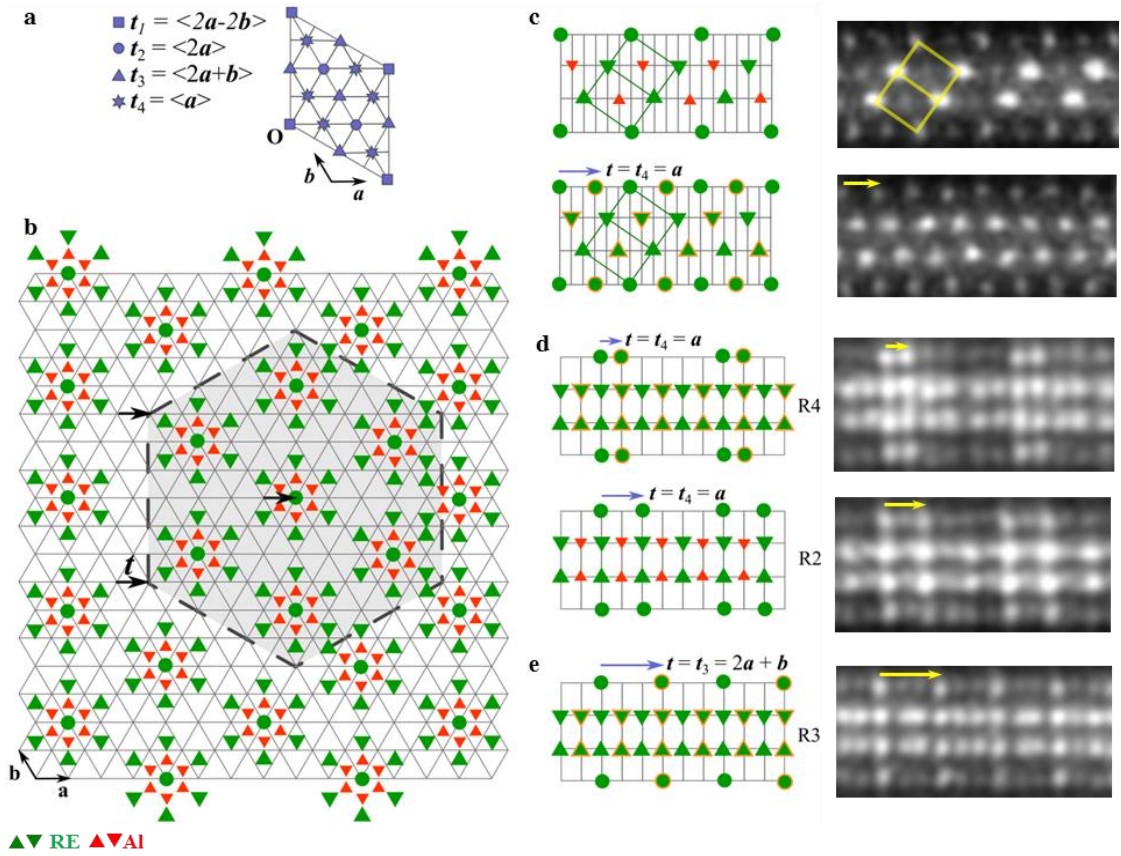

Figure 2. Schematic diagram of possible translation vectors for domain structures and different views of the domain structures. a) Four non-equivalent positions ($t_1$-$t_4$) in A layer for generating the domain structure, where $t_1$ positions are the original positions for ordered clusters. Relative to the origin $O$, the positions could be denoted as $t_1$ = <2$a$-2$b$> = <2-200>$_α$, $t_2$ = <2$a$> = <4-2-20>$_α$/3, $t_3$ = <2$a$+$b$> = <10-10>$_α$, $t_4$ = <$a$> = <2-1-10>$_α$/3, where $a$ and $b$ is the base vectors in the basal plane in Mg ($α$) matrix. b) A region marked by dashed line in Fig. 1a is translated by $t$ = $t_4$ = <2-1-10>$_α$/3. c) Two possible structures when Fig. b is viewed along different <11-20>$_α$, and a HAADF-STEM image is attached at right side for reference. d) Two possible structures when Fig. b is viewed along different <10-10>$_α$. These ordered patterns are named as "R4" and "R2" type structures in Ref. [1], respectively. e) The <10-10>$_α$ view of "R3" type ordered pattern for domain structure located at $t_3$ positions.

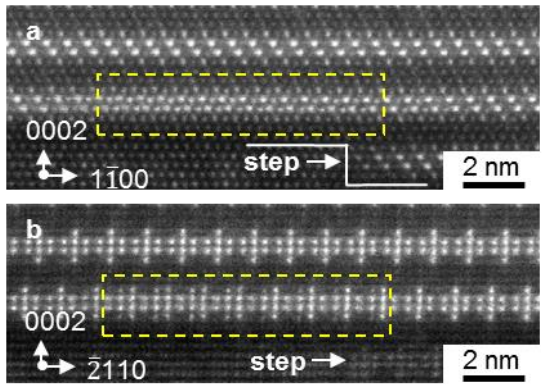

Figure 3. Ordered patterns viewed at different zone axis by HAADF-STEM [6]. a) [11-20]$_\alpha$ zone axis view. b) corresponding [01-10]$_\alpha$ zone axis view, where the 'R2' type clusters are marked by a yellow box.

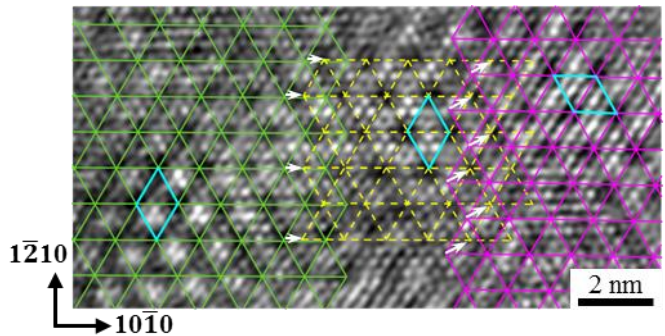

Figure 4. Direct observation of domain structures along [0001]$_\alpha$ by HAADF-STEM. The possible distribution of clusters is indicated by grids with different color, and the mismatching between the grid lattices indicates the existence of domain structure.